
\input phyzzx
\singlespace
\twelvepoint
\REF\bv{N. Berkovits and C. Vafa, {\sl On the Uniqueness of String Theory},
preprint HUTP-93/A031, KCL-TH-93-13, hep-th/9310170 (1993).}
\REF\va{D.V. Volkov and V.P. Akulov, {\sl JETP Lett.} {\bf 16}, 438 (1972).}
\REF\bo{N. Berkovits and N. Ohta, {\sl Embeddings for Non-Critical
Superstrings}, preprint KCL-TH-94-6, OU-HET 189, hep-th/9405144 (1994).}
\REF\col{S. Coleman, {\sl Dilatation}, in {\it Properties of the Fundamental
Interactions}, Editrice Compositori, Bologna, 1973.}
\REF\dz{D.V. Volkov and V.A. Soroka, {\sl JETP Lett.} {\bf 18}, 312 (1973);
\nextline
S. Deser and B. Zumino, {\sl Phys. Rev. Lett.} {\bf 38}, 1433 (1977).}
\REF\bfw{N. Berkovits, M. Freeman and P. West,
{\sl Phys. Lett.} {\bf B325}, 63 (1994).}
\REF\skt{H. Kunitomo, M. Sakaguchi and A. Tokura,
{\sl A Universal $w$\ String Theory},
preprint OU-HET 187, hep-th/9403086 (1994).}
\REF\bop{F. Bastianelli, N. Ohta and J.L. Petersen,
{\sl A Hierarchy of Superstrings},
preprint NBI-HE-94-20, hep-th/9403150 (1994).}
\REF\hie{H. Kunitomo, M. Sakaguchi and A. Tokura,
{\sl A Hierarchy of Super $w$\ Strings},
preprint in preparation.}
%
%
\pubnum={OU-HET 196}
\titlepage
\title{On the Nonlinear Realization of the Superconformal Symmetry}
\author{Hiroshi KUNITOMO\foot{e-mail: kunitomo@oskth.kek.jp}}
\address{\null\hskip-8mm
Department of Physics, Osaka University, \break
Toyonaka, Osaka 560, JAPAN}
\abstract{
The nonlinear realization of the superconformal symmetry in two dimensions is
considered. The superconformal symmetry is realized by means of dimension
$-1/2$\ Nambu-Goldstone fermion $\xi$\ and its dimension 3/2 conjugate $\eta$.
A matter coupling of these Nambu-Goldstone fermions reproduce the realization
found by Berkovits and Vafa to show the equivalence between the bosonic string
and a $N=1$\ superstring. This indicates that this equivalence is a result of
the super Higgs mechanism as a two dimensional field theory, in which the NG
fermions become unphysical.
}
\endpage

Recently, Berkovits and Vafa showed that the bosonic strings may be viewed as a
particular class of vacuua for $N=1$\ superstrings\refmark{\bv}. One of their
key observation is that there is a realization of superconformal symmetry by
means of fermions $(\eta,\xi)$\ with dimensions $(3/2,-1/2)$\  and matter
stress tensor $T_m$. In this realization, the fermion $\xi$\ is transformed
inhomogeneously and thus plays a role of the Nambu-Goldstone (NG) fermion.

In this letter, we rederive their result from the viewpoint of the conventional
nonlinear realization of the supersymmetry\refmark{\va}. The superconformal
symmetry may be nonlinearly realized by NG fermion pair $(\eta,\xi)$. Its
general matter couplings reproduce noncritical embedding obtained by Berkovits
and Ohta\refmark{\bo}, which is including the above realization as a critical
case.

We restrict our attention to the holomorphic sector of the superconformal
symmetry. For studying the nonlinear realization, we consider the
superconformal symmetry as coordinate transformations of holomorphic
superspace. The infinitesimal superconformal symmetry is represented by
$$
\eqalignno{
z'&=z-f(z)+\theta\epsilon (z),\cr
\theta '&=\theta-{1\over 2}\partial f(z)\theta +\epsilon (z),
&\eqname\sct\cr
}
$$
where $f(z)\ (\epsilon (z))$\ is a infinitesimal parameter of the (super)
conformal transformation. We consider such a nonlinear realization that the
superconformal symmetry breaks down to the conformal symmetry. The nonlinear
transformation law of the NG fermion $\xi$\ is obtained by replacing $\theta$\
with NG fermion $\xi$\ in \sct\foot{
Note that we need only one NG fermion while the superconformal symmetry has
infinite dimensions. This seems to be a common feature when broken symmetries
are generated by moments of {\it one} current\refmark{\col}.}:
$$
\eqalignno{
\xi '(z',\bar z)&=\xi (z,\bar z)-{1\over 2}\partial f(z)\xi (z,\bar z)
+\epsilon (z),
&\eqname\ngfNG\cr
z'&=z-f(z)+\xi (z,\bar z)\epsilon (z).
&\eqname\ngf\cr
}
$$
Thus the infinitesimal transformation of $\xi$\ is given by
$$
\eqalignno{
\delta (f,\epsilon)\xi (z,\bar z)&=\xi '(z,\bar z)-\xi (z,\bar z)\cr
&=\epsilon (z)+\epsilon (z)\xi(z,\bar z)\partial\xi (z,\bar z)
-{1\over 2}\partial f(z)\xi (z,\bar z)+f(z)\partial\xi (z,\bar z),
&\eqname\tl\cr
}
$$
where we note that the NG field $\xi$\ is transformed as a dimension $-1/2$\
field under the conformal transformation.
One can show that this transformation satisfies the required commutation
relations:
$$
\eqalignno{
[\delta (f_1,\epsilon_1),\delta (f_2,\epsilon_2)]\xi&=
\delta (f_{12},\epsilon_{12})\xi,
&\eqname\com\cr
}
$$
where
$$
\eqalignno{
f_{12}&=f_1\partial f_2-f_2\partial f_1-2\epsilon_1\epsilon_2,\cr
\epsilon_{12}&=-{1\over 2}\partial f_1\epsilon_2+f_1\partial\epsilon_2
-(1\leftrightarrow 2).
}
$$

Since $\xi$\ have negative dimensions $-1/2$, we cannot obtain nontrivial
conformal invariant theory by using the NG fermion $\xi$\ alone. This is a
characteristic feature of the spontaneous breaking of the superconformal
symmetry. Infinite dimensional unbroken symmetry, conformal symmetry, requires
another degree of freedom than the NG fermion $\xi$. The simplest way to
construct it is introducing a dimension $3/2$\ fermion $\eta$\ and taking the
action:
$$
\eqalignno{
S&=\int d^2z\eta\bar\partial\xi ,
&\eqname\act\cr
}
$$
which implies
$$
\eqalignno{
\eta (z)\xi(w)&\sim {1\over z-w}.
&\eqname\ff\cr
}
$$
The generators of transformations \tl\ are written by using these fields as
$$
\eqalignno{
T_f&=\oint {dz\over 2\pi i}f(z)T^0_{NG}(z),\qquad\qquad
G_{\epsilon}=\oint {dz\over 2\pi i}\epsilon (z)G^0_{NG}(z),
&\eqname\gen\cr
}
$$
where
$$
\eqalignno{
T^0_{NG}(z)&=-{3\over 2}\eta\partial\xi-{1\over 2}\partial\eta\xi,\cr
G^0_{NG}(z)&=\eta+\xi\partial\xi\eta.
&\eqname\zero\cr
}
$$
These also generate such superconformal transformations on $\eta (z)$\ under
that the free field action \act\ is invariant.
By keeping only single contraction, which is equivalent to taking the Poisson
bracket, these generators \gen\ satisfy the classical (centerless)
superconformal algebra. To satisfy exact OPE relations, we must replace
classical currents \zero\ with the quantum ones which have some quantum
correction terms. The final form of the nonlinear realization of the
superconformal currents are given by
$$
\eqalignno{
T_{NG}&=-{3\over 2}\eta\partial\xi-{1\over 2}\partial\eta\xi
+{1\over 2}\partial^2(\xi\partial\xi),\cr
G_{NG}&=\eta+\eta\xi\partial\xi
-{11\over 6}\partial^2\xi+{13\over 12}\xi\partial\xi\partial^2\xi.
&\eqname\NG\cr
}
$$
These satisfy the OPE of superconformal algebra with $c_{NG}=-11$.

We next consider general matter coupling, where matter fields are linear
representations of the conformal symmetry. We denotes a current of this matter
conformal symmetry (stress tensor) as $T_m$. The infinitesimal transformations
of the matter fields under the superconformal symmetry is given by taking \ngf\
as NG field dependent conformal transformations. For instance, the stress
tensor itself is transformed as
$$
\eqalignno{
T_m'(z')&=T(z)-2\partial(z'-z)T_m(z),\cr
\delta (f,\epsilon)T_m(z)&=T_m'(z)-T_m(z)\cr
&=2\partial fT_m+f\partial T_m+2\partial\epsilon\xi T_m+2\epsilon\partial\xi
T_m
+\epsilon\xi\partial T_m.
&\eqname\matf\cr
}
$$
Therefore the classical superconformal currents of the coupled system can be
written by using the stress tensor $T_m$\ as
$$
\eqalignno{
T^0&=T^0_{NG}+T_m,\qquad\qquad
G^0=G^0_{NG}+\xi T_m.
&\eqname\cop\cr
}
$$
Including quantum corrections, the superconformal currents are written by
$$
\eqalignno{
T(z)&=T_{NG}+T_m,\cr
G(z)&=G_{NG}+\xi T_m+{c_m\over 6}\partial^2\xi
-{c_m\over 24}\xi\partial\xi\partial^2\xi\cr
&=\eta+\eta\xi\partial\xi+\xi T_m
+{c_m-11\over 6}\partial^2\xi
+{26-c_m\over 24}\xi\partial\xi\partial^2\xi,
&\eqname\tot\cr
}
$$
where $c_m$\ denotes the central charge of the matter stress tensor $T_m$.
Generators \tot\ coincide with those given in [\bo] and satisfy the
superconformal algebra with $c=c_m-11$, which including the realization
obtained in [\bv] as a critical $c_m=26$\ case.

In this letter, we have reproduced the realization of superconformal algebra
found in [\bv,\bo] from the viewpoint of the standard nonlinear realization.
This indicates that the $N=1$\ superstring studied in [\bv] is in a
spontaneously broken phase of the superconformal symmetry as a two dimensional
field theory. When we consider the superstrings, however, the supersymmetry is
a local symmetry and thus the NG fermion becomes unphysical owing to the super
Higgs mechanism\refmark{\dz}. The physical states, therefore, coincide with
those of the bosonic string.

The results can be generalized to the other embedding\refmark{\bfw -\hie} in
principle. However, we do not have general procedure yet for obtaining quantum
correction which is infinite series in some cases\refmark{\bo}.

\endpage
\centerline{{\bf Acknowledgements}}

The author would like to thank Taichiro Kugo, Nobuyoshi Ohta, Makoto Sakaguchi
and Akira Tokura for discussions.

\refout
\bye